# GESTALT PHENOMENON IN MUSIC?
# A NEUROCOGNITIVE PHYSICS STUDY WITH EEG

Shankha Sanyal[1,2], Archi Banerjee[1,2], Souparno Roy[1,2*], Sourya Sengupta[3], Sayan Biswas[3], Sayan Nag[3], Ranjan Sengupta[1] and Dipak Ghosh[1]

[1]Sir C.V. Raman Centre for Physics and Music, Jadavpur University

[2]Department of Physics, Jadavpur University

[3]Department of Electrical Engineering, Jadavpur University

* Corresponding Author

**ABSTRACT**

*The term "gestalt" has been widely used in the field of psychology which defined the perception of human mind to group any object not in part but as a 'unified' whole. Music in general is polytonic - i.e. a combination of a number of pure tones (frequencies) mixed together in a manner that sounds harmonius. The study of human brain response due to different frequency groups of acoustic signal can give us an excellent insight regarding the neural and functional architecture of brain functions. In this work we have tried to analyze the effect of different frequency bands of music on the various frequency rhythms of human brain obtained from EEG data of 5 participants. Four (4) widely popular Rabindrasangeet clips were subjected to Wavelet Transform method for extracting five resonant frequency bands from the original music signal. These resonant frequency bands were presented to the subjects as auditory stimulus and EEG signals recorded simultaneously in 19 different locations of the brain. The recorded EEG signals were noise cleaned and subjected to Multifractal Detrended Fluctuation Analysis (MFDFA) technique on the alpha, theta and gamma frequency range. Thus, we obtained the complexity values (in the form of multifractal spectral width) in alpha, theta and gamma EEG rhythms corresponding to different frequency bands of music. We obtain frequency specific arousal based response in different lobes of brain as well as in specific EEG bands corresponding to musical stimuli. This revelation can be of immense importance when it comes to the field of cognitive music therapy.*

**Keywords**: Gestalt in music, EEG, alpha, theta and gamma response, Multifractal analysis

*thesouparnoroy@gmail.com

# INTRODUCTION

How does human being perceive and recognize a musical sound? Is there any specific frequency region of music which makes a particular song identifiable? If so, what is the corresponding brain response to that particular frequency band? These are a few questions which we try to venture into with the help of this study. Although a number of previous works have studied that the human ear has an enhanced sensitivity in the frequency range of 2 kHz to 5 kHz, i.e. it reacts most to the frequencies of these ranges [1], till date, very few works [2, 3] analyze the response of human brain to different bands of musical frequencies. Our work tries to visualize whether the sensitivity of human ear (and also human brain) to different frequencies of song (including lyrics and melodic content) is same as that found earlier experiments or is there some different element which makes it distinct from earlier approaches.

According to a researcher at Cambridge University [4], it doesn't matter in what order the letters in a word are, the only important thing is that the first and last letter be at the right place. The rest can be a total mess and you can still read it without problem. This is because the human mind does not read every letter by itself but the word as a whole. The whole idea of gestalt phenomenon in music originated from this experiment which is carried out mainly in the visual domain; we thought of reproducing the same type of experiment in the auditory domain which led us to developing the protocol of this particular experiment. In case of musical stimulus, frequency and time can be considered as the parameters which play the same role as the arrangement of letters in a word. If we can vary the pitch profile of a particular song or can randomize the time gap between the throw of certain words of a song, there must exist a threshold value where the song becomes non-perceptible to the listener. With this work, we delve into the study of one of this parameter - i.e. several sub-bands of frequency were extracted and their respective effect on human brain were analyzed

*Gestalt* is a psychology term which means "unified whole". It refers to theories of visual perception developed by German psychologists in the 1920s. These theories attempt to describe how people tend to organize visual elements into groups or *unified wholes* when certain principles are applied [5]. The various principles which make use of gestalt psychology are similarity, continuation, closure and proximity. A number of experiments in the visual domain tries to explore the paradigm of gestalt psychology using ambiguous figures as visual stimuli.

There is a group of research work which deals with the creative manifestations of gestalt therapy [6, 7]; which begin with the application of the principles of Gestalt theory, such as figure/ground, the principles of good Gestalt, Prägnanz and closure, as well as viewing perception as an active process. Creativity in Gestalt theory means venturing beyond self-expression and entering the dynamics of the productive interchange within the therapeutic relationship, and thus creation of a space which is not ventured earlier. A formal definition for creative thinking, as ascertained by Guilford [8] is "conceptual redefinition" or "the ability to redefine or reorganize objects of thought". But all these works look into gestalt therapy mainly from the view of art and craft and not any scientific phenomenon. In this work, we envisaged to study gestalt phenomenon in a new light of sound scientific principles and the inherent brain dynamics associated with it.

Although the main focus of gestalt theory has been on the visual domain, a number of works have been done in the auditory domain as well [9-11], using the various principles of gestalt theory viz. proximity, similarity closure etc. If seen loosely, visual and auditory perceptions are poles apart from one another; one is spatial perception, the other temporal. However it is easy to see how each contains elements of the other - visual perception also changes over time, just like musical/auditory perception; when we look at moving or changing forms, even when we see a static image our eyes move across it in meaningful patterns. No study, till date explores the brain response corresponding to a musical clip which has been doctored and grouped keeping in mind the principles of gestalt theory.

Each type of music has its own frequency, which can either resonate or be in conflict with the body's rhythms (heart rate). Studying EEG dynamics typically relies on the calculation of temporal and/or spectral dynamics from signals recorded directly from the scalp. Each frequency band of the EEG rhythm relates to specific functions of the brain. EEG rhythms are classified into five basic types: i) delta (δ) 0.5-4Hz, (ii) Theta (θ) 4–8 Hz, (iii) alpha (α) 8-13Hz, (iv) beta (β) 13-30 Hz and (v) gamma (γ) 30-50 Hz. It has been observed that pleasant music produces a decrease in the alpha power at the left frontal lobe and unpleasant music produces decrease in the alpha power at the right frontal lobe [12-14]. Also, activity in the alpha frequency band has been found to be negatively related to the activity of the cortex, such that larger alpha frequency values are related to lower activity in the cortical areas of the brain, while lower alpha frequencies are associated with higher activity in the cortical areas [15, 16]. The Fm theta power was positively correlated not only with scores of internalized attention but also with subjective scores of the pleasantness of the emotional experience. Furthermore, two studies on the relationship between Fm theta and anxiety reported negative correlations between Fm theta during mental tasks and anxiety measures [16, 17]. It has also been shown that pleasant music would elicit an increase of Fm theta power [14, 18]. While listening to music, degrees of the gamma band synchrony over distributed cortical areas were found to be significantly higher in musicians than non musicians [19]. The gamma band EEG distributed over different areas of brain while listening to music can be represented by a universal scaling which is reduced during resting condition as well as when listening to texts [20].

The scalp EEG arises from the interactions of a large number of neurons whose interactions generally nonlinear and thus they can generate fluctuations that are not best described by linear decomposition. DFA [21] has been applied to EEG signals to identify music induced emotions in a number of studies [22-24]. Gao et.al [22] related emotional intensity with the scaling exponent, while a recent study [24] relate the variation of alpha scaling exponent generated from DFA technique with the retention of musical emotions – an evidence of hysteresis in human brain. But DFA has its own limitations. Many geophysical signals as well as biosignals do not exhibit monofractal scaling behavior, which can be accounted for by a single scaling exponent [25, 26], therefore different scaling exponents are required for different parts of the series [27]. Consequently a multifractal analysis should be applied. The Multifractal Detrended Fluctuation Analysis (MFDFA) technique was first conceived by [28] as a generalization of the standard DFA. MFDFA has been applied successfully to study multifractal scaling behavior of various non-stationary time series [29-31]. EEG signals are essentially multifractals as they consist of segments with large variations as well as segments with very small variations, hence when applied to the alpha and theta EEG rhythms, the multifractal spectral width will be an indicator of emotional arousal corresponding to particular clip. In case of music induced emotions, a recent study [32] used the multifractal spectral width as an indicator to assess emotional arousal corresponding to the simplest musical stimuli – a tanpura drone.

In this work we look forward to study brain response to different frequency bands of 4 musical clips – i.e. 4 pre-recorded Tagore songs, sung by a renowned artist without any accompaniment. Music in general is polytonic, i.e. a number of pure tones mixed together in such a way that it sounds harmonius; musical sound produced by human voice in this way is also quasi-periodic, albeit highly complex in nature. To avoid any variation arising due to the change of timbral parameters, recordings from the same singer were taken for our experiment. From the complete recording, approximately 20 second clips were clipped and segregated into five frequency bands viz. Band 1 (50Hz to 1 kHz), Band 2 (1 kHz to 2 kHz), Band 3 (2 kHz to 3 kHz), Band 4 (3 kHz to 4 kHz) and Band 5 (4 kHz and above) using band pass filter in Fast Fourier Transform (FFT) technique. The frequency bands were so chosen that each band contains two or more overtones/resonant frequency bands; which essentially contains the cue for identification of a particular song. Next a human response data was collected from about 50 participants, where the respondents were asked to mark the band in which they could not identify the song. From the obtained classification table (**Table 1**), it was seen that maximum non-recognition is being seen in the $4^{th}$ and $5^{th}$ band, while in spite of the removal of fundamental frequency and two/three higher harmonics, the respondents can clearly recognize the song in the $2^{nd}$ and $3^{rd}$ band. From this observation it is clear that human mind can recognize musical timbre till around 3 kHz even without the presence of fundamental frequency, while a switch occurs above that, which leads to its non-recognition. We sought to understand the EEG brain response of this switch, where the human mind is unable to process the frequency bands. For this, a pool of 5 subjects was randomly chosen from the pool of 50 respondents who participated in the human response study and EEG experiment was performed on them using the same protocol in which psychological response was taken. Next, 10 electrodes were chosen (F3, F4, F7, F8, T3, T4, T5, T6, O1 and O2) from different locations of the brain whose modalities match with our work, and time series data were extracted from each one of them. The time series data obtained from each of the 10 electrodes were separated into alpha, theta and gamma frequency rhythms for each of the experimental condition and analyzed with the help of MFDFA technique. The multifractal spectral width obtained for each of the experimental condition acts as a parameter with which we can identify the arousal based activities in different lobes of brain. From the results obtained we see there is a definite switch in the alpha, theta and gamma complexities in different lobes in response to $4^{th}$ and $5^{th}$ band (where there is non-recognition), with the response most significant in the temporal lobe and gamma band. With this work, we try to venture into a hitherto unexplored domain of human brain response to different frequency bands of music, which may be a cue to study the gestalt principles in the auditory domain. The results and implications are discussed in detail in the following sections.

## MATERIALS AND METHODS
### A. SUBJECTS SUMMARY
Participants for listening test data were recruited mostly through word of mouth and social media platforms. All the listening test data were collected at the Sir C.V. Raman Centre for Physics and Music, Jadavpur University over a period of 2 months. In this study, response from 50 participants ($F = 17$, $M = 13$) were considered for analysis presented herewith, who participated voluntarily. The subjects chosen had no formal musical training.

5 (M=3, F=2) musically untrained adults chosen randomly from the pool created from listening test data who voluntarily participated in this study. The average age was 23 years (SD= 2.35 years) and average body weight was 60 kg. Informed consent was obtained from each subject according to the guidelines of the Ethical Committee of Jadavpur University. All experiments were performed at the Sir C.V. Raman Centre for Physics and Music, Jadavpur University, Kolkata. The experiment was conducted in the afternoon with a normal diet in a normally conditioned room sitting on a comfortable chair and performed as per the guidelines of the Institutional Ethics Committee of SSN College for Human volunteer research.

### B. EXPERIMENTAL DETAILS

For the listening test, a template like the one given in Fig. 1 were made for each of the sample and presented to the respondents. The template consists of the clips in the following order:
Original Clip =>  Band 3 (2 kHz to 3 kHz) (Part 1) => Band 2 (1 kHz to 2 kHz) (Part 2) => Band 5   (4kHz to 5kHz) (Part 3) => Band 4 (3kHz to 4 kHz) (Part 4) => Band 1 (50Hz to 1 kHz) (Part 5). A resting time of about 5 seconds were given between each successive clip. All the clips were normalized to keep the amplitude constant.

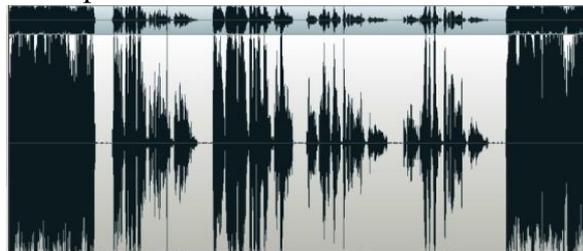

**Fig. 1**: The template of music clip containing different frequency bands played randomly

Approximately 20 second clips from four popular Tagore songs sung by a renowned singer (without any accompaniment) were taken for our analysis, with the original clip being played first, followed by the jumbled order of clips. An Instruction Sheet along with a response form like the one given below (**Fig. 2**) was given to the respondents where they were asked to mark the parts where they could not recognize the song.

|        | Part 1 | Part 2 | Part 3 | Part 4 | Part 5 |
|--------|--------|--------|--------|--------|--------|
| Clip 1 |        |        |        |        |        |
| Clip 2 |        |        |        |        |        |
| Clip 3 |        |        |        |        |        |
| Clip 4 |        |        |        |        |        |

**Fig. 2**: Response Sheet to identify the non-recognition of each song

During the EEG acquisition period, the 5 subjects were made to listen to the same clips in the same order as in the listening test. Each experimental condition lasted for around 12 min. Each song clip of 20 second was followed by a resting period of 5 seconds during which no music was played. Subjects were asked to keep their eyes closed and to sit calmly during each condition. First, the baseline (i.e. a resting condition) was recorded for each subject before the start of the experiment with 1 min of 'no music' condition.

The music was presented with the computer-sound system (Logitech R _ Z-4 speakers) with very high S/N ratio was used in the measurement room for giving music input to the subjects ca. 120 cm behind the head of the subjects with a volume of 45–60 dB. The EEG experiment was conducted in the afternoon (around 2 PM) in a room with the volunteers sitting in a comfortable chair.

## C. EXPERIMENTAL PROTOCOL

Since the objective of this study was to analyze the effect of cross-cultural contemporary instrumental music on brain activity during the normal relaxing condition, the frontal, temporal and occipital lobes were selected for the study. EEG was done to record the brain-electrical response of 20 subjects. Each subject was prepared with an EEG recording cap with 19 electrodes (Ag/AgCl sintered ring electrodes) placed in the international 10/20 system. **Figure 3** depicts the positions of the electrodes. Impedances were checked below 50 kΩ. The EEG recording system (Recorders and Medicare Systems) was operated at 256 samples/s recording on customized software of RMS. The data was band-pass-filtered between 0.5 and 35Hz to remove DC drifts and suppress the 50Hz power line interference. The ear electrodes A1 and A2 linked together have been used as the reference electrodes. The forehead electrode, FPz has been used as the ground electrode. Each subject was seated comfortably in a relaxed condition in a chair in a shielded measurement cabin. They were also asked to close their eyes. After initialization, a 10 minutes recording period was started, and the following protocol was followed:

1. 60 seconds No Music (Resting Condition)
2. 20 seconds Clip 1 (Original)
3. 5 seconds No Music
4. 20 seconds Clip 1 Band 3 (Part 1)
5. 5 seconds No Music
6. 20 seconds Clip 1 Band 2 (Part 2)
7. 5 seconds No Music
8. 20 seconds Clip 1 Band 5 (Part 3)
9. 5 seconds No Music
10. 20 seconds Clip 1 Band 4 (Part 4)
11. 5 seconds No Music
12. 20 seconds Clip 1 Band 1 (Part 1)
13. 30 second Resting period

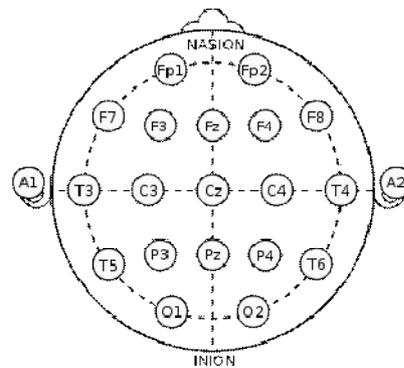

**Fig. 3** The position of electrodes according to the 10-20 international system

The same protocol was repeated for Clips 2, 3 and 4 with a 30 second resting period in between each clip. Markers were set at start, signal onset/offset, and at the end of the recording.

## METHODOLOGY

We have obtained noise free EEG data for all the electrodes using the EMD technique as in [32] and used this data for further analysis and classification of acoustic stimuli induced EEG features. The amplitude envelope of the alpha (8-13 Hz), theta (4-7 Hz) and gamma (13-30 Hz) frequency ranges were obtained using wavelet transform technique. Data was extracted for these electrodes according to the time period given in the Experimental protocol section i.e. for Experimental conditions 1 to 13.

### A. Wavelet Transform:

Wavelet transform (WT) forms a general mathematical tool for time-scale signal analysis and decomposition of EEG as well as music signal. We have used WT technique to decompose the EEG and complete music signal into various frequency bands. The DWT [33] analyzes the signal at different frequency bands with different resolutions by decomposing the signal into a coarse approximation and obtains detailed information. DWT generally employs two sets of functions, called the scaling functions and wavelet functions, associated with low pass and high pass filters, respectively. The decomposition of the signal into different frequency bands is done by successive high pass and low pass filtering of the time domain signal. In this way time series data of alpha, gamma and theta EEG waves as well as desired frequency bands of music signal were obtained corresponding to each experimental condition. On the obtained EEG time series data, MFDFA analysis was performed.

**B. Method of multifractal analysis of EEG signals:**
The analysis of the alpha and theta EEG signals are done using MATLAB [34] and for each step an equivalent mathematical representation is given which is taken from the prescription of Kantelhardt et al [28].
The complete procedure is divided into the following steps:
*Step 1:* The whole length of the signal is divided into Ns no of segments consisting of certain no. of samples. For s as sample size and N the total length of the signal the segments are
$$Ns = int\left(\frac{N}{s}\right)$$
*Step 2:* The local RMS variation for any sample size s is the function F(s,v). This function can be written as follows:
$$F^2(s,v) = \frac{1}{s}\sum_{i=1}^{s}\{Y[(v-1)s+i] - y_v(i)\}^2$$
*Step 3:* The q-order overall RMS variation for various scale sizes can be obtained by the use of following equation
$$Fq(s) = \left\{\frac{1}{Ns}\sum_{v=1}^{Ns}[F^2(s,v)]^{\frac{q}{2}}\right\}^{\left(\frac{1}{q}\right)}$$
*Step 4:* The scaling behaviour of the fluctuation function is obtained by drawing the log-log plot of $F_q(s)$ vs. s for each value of q.
$$F_q(s) \sim s^{h(q)}$$
h (q) is called the generalized Hurst exponent. The Hurst exponent is measure of self-similarity and correlation properties of time series produced by fractal A monofractal time series is characterized by unique h(q) for all values of q.

The generalized Hurst exponent h(q) of MF-DFA is related to the classical scaling exponent τ(q) by the relation
$$\tau(q) = qh(q) - 1$$
A monofractal series with long range correlation is characterized by linearly dependent q order exponent τ(q) with a single Hurst exponent H. Multifractal signal on the other hand, possess multiple Hurst exponent and in this case, τ(q) depends non-linearly on q [38]. The singularity spectrum f(α) is related to h(q) by
$$\alpha = h(q) + qh'(q)$$
$$f(\alpha) = q[\alpha - h(q)] + 1$$
where α is the singularity strength and f(α) specifies the dimension of subset series that is characterized by α. The multifractal spectrum is capable of providing information about relative importance of various fractal exponents in the series e.g., the width of the spectrum denotes range of exponents. A quantitative characterization of the spectra may be obtained by least square fitting it to a quadratic function [35] around the position of maximum $α_0$,

$$f(α) = A(α − α_0)^2 + B(α − α_0) + C$$

where C is an additive constant $C = f(α_0) = 1$. B indicates the asymmetry of the spectrum. It is zero for a symmetric spectrum. The width of the spectrum can be obtained by extrapolating the fitted curve to zero.
Width W is defined as,

$$W = α_1 − α_2,$$
with $f(α_1) = f(α_2) = 0$.

The width of the spectrum gives a measure of the multifractality of the spectrum. Greater is the value of the width W greater will be the multifractality of the spectrum. For a monofractal time series, the width will be zero as h(q) is independent of q. The origin of multifractality in a EEG time series can be verified by randomly shuffling the original time series data [32, 36].

## RESULTS AND DISCUSSIONS:

From the results of human response data, a percentage chart like the one given in Table 1 is plotted where the recognition percentage for each frequency band is given for a population of 50 people:

**Table 1**: Amount of Non-Recognition of Song from a listening test of 50 informants (in percentage)

|        | Band 1 | Band 2 | Band 3 | Band 4 | Band 5 |
|--------|--------|--------|--------|--------|--------|
| Clip 1 | 0      | 0      | 15     | 78     | 100    |
| Clip 2 | 0      | 0      | 12     | 87     | 95     |
| Clip 3 | 0      | 0      | 5      | 97     | 100    |
| Clip 4 | 0      | 0      | 20     | 89     | 95     |

It is seen for most of the clips, there is a definite switch in human perception above Band 3 i.e. above 3 kHz frequency; the human ear is not able to recognize the song from somewhere above this frequency range; which follows that in Band 5 (which contains frequencies from 4 kHz and above) almost all the participants were unable to recognize the song. In the following sections we will try to discuss the brain correlates associated with this switch in perception. From the human response study, thus we get an important cue that up to 3 kHz frequency human brain follows the principle of closure and is able to perceive the entire song even when the fundamental along with a few more harmonics have been cut off. But above that, mostly there is non-recognition for that particular song except for certain percentage of listeners who are able to perceive the song even above that from the melodic cues attached to a specific song.

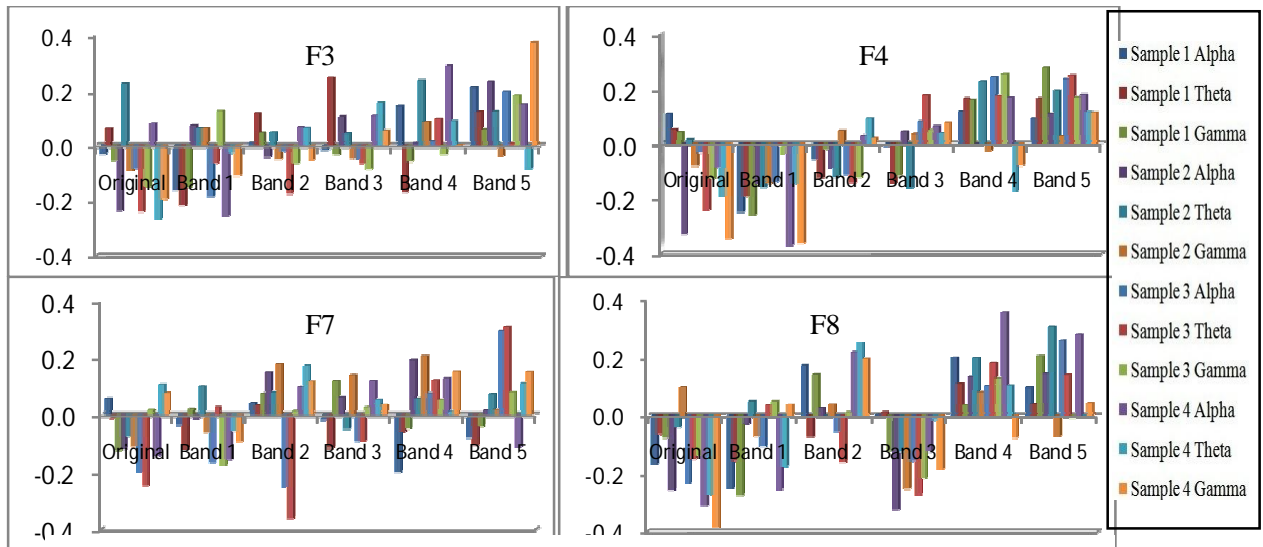

**Fig. 3a- 3d** : Variation of multifractal width for frontal F3, F4, F7 and F8 electrodes

The amount of multifractality and hence the complexity can be determined quantitatively in each of the electrode from the width of the multifractal spectrum [**f(α) vs α**]. The values of multifractal widths have been averaged for the 5 persons and the variation of complexities from the resting (no music condition) have been computed for all the experimental conditions. **Fig. 3a-d** plot the variation in alpha, theta and gamma complexities in response to the various frequency bands of 4 music graphically for each of the 10 electrodes chosen for our study.

A general glance into the figures help us to identify that there is a characterestic switch from Band 3 to Band 4, in the complexities corresponding to each of the frequency band in all the frontal electrodes. While for all the frequency bands upto 3, there is a general decrease in alpha, theta and gamma complexities, a sudden spike in the complexities is seen corresponding to Band 4 and Band 5, which are the regions of non-recognition as is verified from the psychological test data. The sudden increase in complexity is very prominent in the alpha and gamma frequency range of right frontal electrodes F4 and F8. The theta frequency range gives somewhat ambiguous results wherein any specific pattern is not observed for the various frequency bands of music given as input. For some songs, specifically 3 and 4, however we find that the alpha and theta complexities taking a jump from Band 3 only; which may be a cue that the respondents had some confusion in perceiving the song from this frequency band also. The variation in multifractal widths for the four temporal electrodes is given in **Fig. 4a-d**.

For the temporal electrodes, the switch from recognition to non-recognition shows the same signature as in the case of frontal electrodes, but since the temporal lobes are generally associated with auditory processing, the manifestation of change is also very strong here. In this case, however we find the response is strongest in the T5 and T6 electrodes, whereby a definite rise in alpha and gamma complexity is seen in Band 4 and Band 5 as compared to their fall in other frequency bands of music. An interesting observation here is that for Sample 4, the alpha complexity decreases for all the temporal electrodes which may be a symbol of recognition of this particular music clip, although theta and gamma complexity registers an increase.

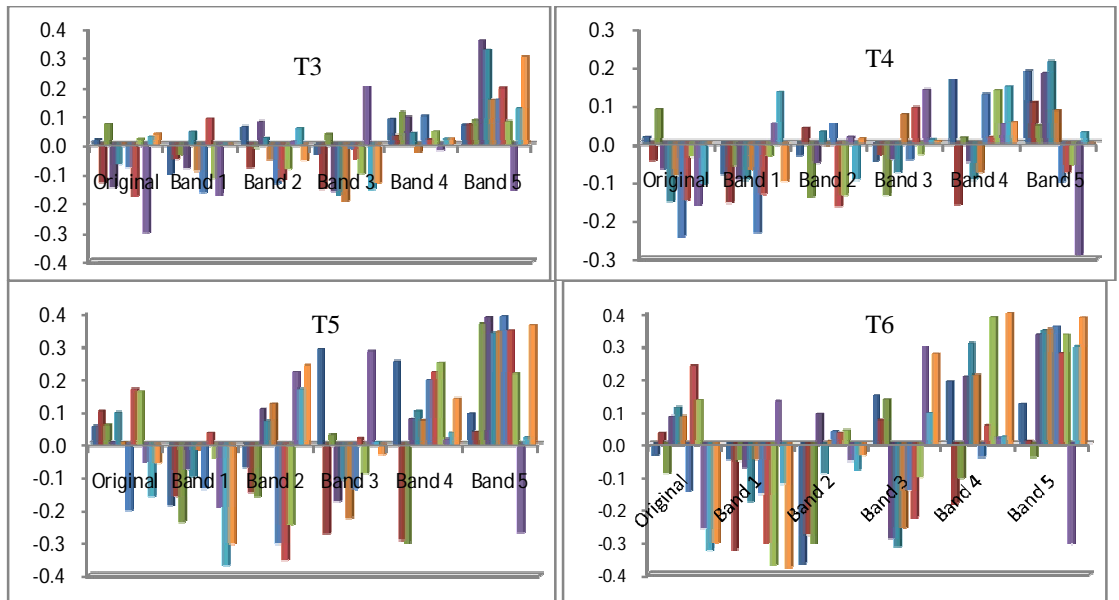

**Fig. 4a- 4d** : Variation of multifractal width for temporal T3, T4, T5 and T6 electrodes

For Sample 1, the theta and gamma complexity reports a decrease under the effect of Frquency Band 4 for all the electrodes, which may be a signature of the respondents being able to recognize the song, while in Band 5 their sudden increase may be due to complete non-recognition. In general, for the first two low frequency bands of music, there is a decrease in complexity for all the EEG frequency bands; while for the last two high frequency bands there is a general increase in complexity for most of the EEG frequency bands in case of the temporal electrodes. The figures **5 a-b** show the change in complexities for the two occipital electrodes O1 and O2.

Although, occipital electrodes are associated with processing of visual imagery, any musical piece is associated with some sort of visual imprint in the minds of the listener and hence, the response of occipital electrodes in the perception of different sections of a musical piece is very important. For both the electrodes, we find that the signature for onset of non recognition, i.e a sudden increase in alpha, theta and gamma multifractal width comes at Band 2 for Sample 4, while for Sample 1 it comes at Band 3. The other two samples behave in a similar manner as is found in the temporal and frontal electrodes, with the onset of non-recognition coming at Band 4. For the occipital electrodes, theta band also plays an important role as a marker to detect the switch from recognition to non-recognition of musical frequency bands. We find the strongest response in the occipital electrode for all the EEG frequency ranges in the $1^{st}$ band of music, i.e. which contains the fundamental frequency as well as 2/3 higher harmonics.

This is fairly reasonable as this part contains all the signature of the original song; hence it is easy for the respondents to form a mental imagery for that particular song, which eventually leads to strong arousal in the occipital electrodes for that particular song. Eventually as more and more frequency bands are cut-off from the musical clips, the occipital lobe fails to perceive the original music and hence we see a decreasing arousal based response in Band 2 and a sudden jump in Band 3/ Band 4 where there is complete non-recognition and hence no corresponding visual imagery.

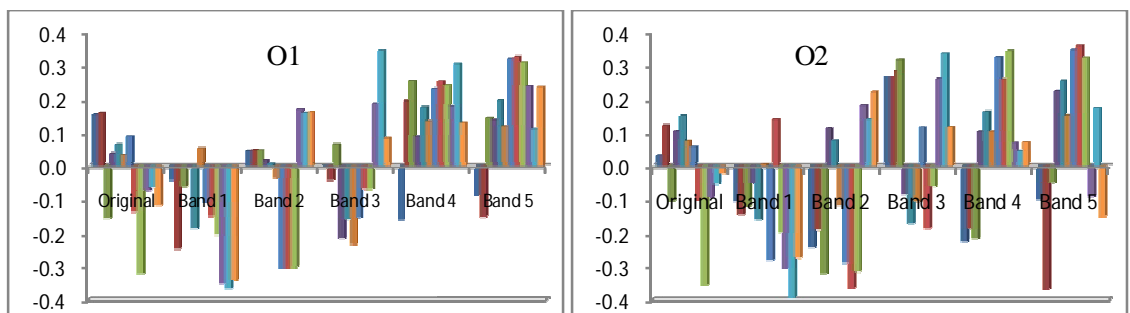
**Fig. 5a- 5b** : Variation of multifractal width for occipital O1 and O2 electrodes

Thus, with the help of this study we have tried to establish a threshold for gestalt phenomenon in music, whereby brain fails to perceive a musical piece above a particular resonant frequency, i.e. the cue for recognition of that particular musical piece is lost, and hence the closure property of gestalt principle fails.

## CONCLUSION

With this work, we tried to venture an unknown horizon of gestalt principle – i.e. is there a cut-off frequency beyond which human mind cannot recognize even a known piece of music. We used a robust non-linear technique for the analysis of EEG data – MFDFA, to identify if there exist such a cut-off frequency for musical piece and what is its manifestation in corresponding EEG response. The study yields the following interesting conclusions:

1. From the human-response data, it is seen that human brain is unable to perceive a musical frequency above a frequency of 3 kHz in general, although there are certain exceptions; but we can safely assume 3 kHz as the switch over frequency while above 4 kHz there is complete non-recognition, i.e. there are almost no musical element left above 4 kHz which will help the human mind to identify the song. The frequency bands below 3 kHz however are easily identified by the participants as the original song, though the fundamental frequency and a number of overtones have been removed from the piece.

2. To find the EEG correlates for the switch, alpha, theta and gamma complexity were studied for 10 electrodes in frontal temporal and occipital region. Statistical analysis showed the arousal based effects were most strong in the right frontal electrodes, F4 and F8 as well as right temporal electrodes T4 and T6; which leads us to the conclusion that perception and recognition based activities are mostly performed in the right frontal and temporal lobe; with the response in T6 and F4 being the strongest in all.

3. We studied the complexity values corresponding to the three frequency ranges of EEG data, with the conclusion that alpha and gamma are the most important markers of gestalt principle in music. The complexity values for both these frequency ranges decrease when the human brain is able to identify a known musical piece, and increase suddenly whenever there is non-recognition of a particular piece. Mostly, we see that switch comes in Band 4 (i.e. for music signals above 3 kHz), and in some exceptional cases, in Band 2 (i.e. between 2 kHz and 3 kHz) also. The theta frequency range, however, plays a significant role in case of the occipital electrodes, where it can be safely used as a marker to distinguish between the two states.

To conclude, we have been able to present a novel evidence of occurrence of Gestalt phenomenon in music. Also, we propose an algorithm with which one can categorize between two states of human mind in response to a well known doctored musical clip which is devoid of certain frequency values. We see that up to a certain frequency band, it can easily perceive the musical piece, while above that range, it becomes unknown. To know the exact value of this cut-off, experiments are going on in the Centre by fine tuning the Band 4 into smaller and smaller groups to yield the exact/range of threshold value(s) where this switch from recognition to non-recognition occurs. This albeit, should not be confused with auditory sensitivity like phenomenon, instead, it hints at a phase transition in self-organized cognitive processes which is a subject of extreme interest in other complex domains. Although we obtained certain variations from Sample to Sample, but those small variations can be neglected as statistical fluctuations, and hence a definite threshold is obtained above which the closure principle of gestalt fails, we are working on a more detailed analysis with a greater number of musical sample and human subjects, which will lead to further refinement and enrichment of the obtained results.

## ACKNOWLEDGEMENT:
One of the authors, AB acknowledges the Department of Science and Technology (DST), Govt. of India for providing (A.20020/11/97-IFD) the DST Inspire Fellowship to pursue this research work. The first author, SS acknowledges the West Bengal State Council of Science and Technology (WBSCST), Govt. of West Bengal for providing the S.N. Bose Research Fellowship Award to pursue this research (193/WBSCST/F/0520/14). SR acknowledges the Department of Science and Technology (DST), Govt. of West Bengal for providing Junior Research Fellowship under Research Project Grant No. [860(Sanc.)/ST/P/S&T/4G-3/2013].